\documentclass[12pt, a4paper]{article}
\usepackage{latexsym,amsfonts,amssymb}
\usepackage{graphicx}
\usepackage{indentfirst}
\usepackage{cite}
\usepackage[cp1251]{inputenc}

\newtheorem{theorem}{Theorem}

\newtheorem{corollary}{Corollary}

\newtheorem{definition}{Definition}

\newtheorem{remark}{Remark}

\begin{document}

\begin{center}
{\Large \bf Lie symmetry of a class of nonlinear boundary value problems with free boundaries}

\medskip

{\bf Roman Cherniha$^{\dag,}$} {\bf and  Sergii Kovalenko$^\dag$}
\\
{\it $^\dag$~Institute of Mathematics, Ukrainian National Academy of Sciences,
\\
3 Tereshchenkivs'ka Street, Kyiv 01601, Ukraine}

\medskip
 E-mail: cherniha@imath.kiev.ua and kovalenko@imath.kiev.ua
\end{center}





\begin{abstract}

A class of (1+1)--dimensional nonlinear boundary value problems
(BVPs), modeling the  process of melting and evaporation of solid
materials, is studied  by means of the classical Lie symmetry
method. New definition of invariance in Lie's sense for  BVP is
presented and applied for the class of BVPs in question.

\end{abstract}

\section{Introduction}

Boundary value problems of the Stefan type  are widely used  in
mathematical  modeling  a huge number of processes, which   arise in
physics,  biology,  and  industry \cite{alex93}, \cite{bri-03},
\cite{crank84}, \cite{ready}, \cite{rub71}. Nevertheless these
processes are very different from   formal point of view, they
have  the common peculiarity, unknown   moving boundaries. Movement
of unknown boundaries is described by  the
 Stefan boundary
conditions. Boundary value problems  with the Stefan conditions are the
main object of this paper.   It is well-known that exact solutions
of BVPs   of the Stefan type  can be derived only in exceptional
cases and the relevant list is very short  at the present time (see
\cite{alex93}, \cite{br-tr02}, \cite{ch93}, \cite{ch-od90},
\cite{voller04} and papers cited therein).

 The  main   idea
is to apply  the classical Lie symmetry method (\cite{b-k},
\cite{fss}--\cite{ovs}) to the case of the Stefan type BVPs. The Lie
symmetry method is the very powerful tool for investigation of
nonlinear partial differential equations (PDEs), notably, for
constructing exact solutions. On the other hand, this method is
 efficient for solving standard BVPs (no  moving boundaries)  in exceptional cases
because the boundary conditions are usually not invariant under any
transformations.

Nevertheless the Stefan type problems are   more complicated objects
than  BVPs with fixed boundaries, we have recently noted
that the Lie symmetry method should be more applicable just for
solving problems with moving boundaries \cite{ch-kov-09}. In fact,
the structure of such boundaries may depend  on invariant
variable(s) and this allows us
 to reduce the given BVP to
one of lower dimensionality. It should be noted that
the rigorous definition of Lie invariance for BVPs and relevant
examples are presented  in \cite{ bl-anco02, b-k}. However,  this definition
does not suit to many realistic BVPs of the Stefan type and should
be modified and generalized.

Let us consider the following  class
of the Stefan type  BVPs used to describe melting and evaporation of materials  in case
that their surface is exposed to a powerful flux of energy
\begin{eqnarray}
& & \frac{\partial}{\partial
x}\left(\lambda_{1}(T_{1})\frac{\partial T_{1}}{\partial x}\right) =
\rho_1(T_1)c_{1}(T_{1}) \frac{\partial
T_{1}}{\partial t},\label{1} \\
& & \frac{\partial}{\partial x}\left(\lambda_{2}(T_{2})
\frac{\partial T_{2}}{\partial x}\right) = \rho_2(T_2)c_{2}(T_{2})
\frac{\partial
T_{2}}{\partial t},\label{2} \\
 & & \qquad S_{1}(t,x) = 0:\ \lambda_{1}(T_{1})
\frac{\partial T_{1}}{\partial x} = \rho_1(T_1) L_v V_1 - Q(t,T_
{1}), \ V_1 = H(t,T_ {1}),\label{3} \\ & & \qquad S_{2}(t,x) = 0: \
\lambda_{2}(T_{m}) \frac{\partial T_{2}}{\partial x} =
\lambda_{1}(T_{m}) \frac{\partial T_{1}}{\partial x} + \rho_2(T_m)
L_m V_2,\ T_{1} = T_{2} = T_{m},\label{4}
\\ & & \qquad x = +\infty: \ T_{2} = T_{0},\label{5}
\end{eqnarray}
where $T_{m}$, $T_{0}$ are the known temperatures of melting and solid
phase of material, respectively; $\lambda_{k}(T_k)$, $c_{k}(T_k)$,
and $\rho_k(T_k)$ are positive thermal conductivities, specific
heat, and densities, respectively; $L_v$, $L_m$ are latent heats of
phase change per unit mass of liquid and solid phases, respectively;
$Q(t,T_1)$ is a function that represents the energy flux being
absorbed by the metal; $H(t,T_ {1})$ is a given function;
$S_{k}(t,x)$ are the phase division boundary surfaces to be found;
$V_k(t,x) = - \frac{\partial S_k}{\partial t} / \frac{\partial
S_k}{\partial x}$ are the phase division boundary velocities;
$T_{k}(t, x)$ are unknown temperature fields; index $k = 1, 2$
corresponds to the liquid and solid phases, respectively.

Here Eqs. (\ref{1})--(\ref{2}) are basic and  describe the heat
transfer process in liquid and solid phases, the boundary conditions
(\ref{3}) present the evaporation dynamics on the surface $S_{1}$, and
the boundary conditions (\ref{4}) are the well-known Stefan
conditions on the surface $S_{2}$ dividing the liquid and solid
phases. Assuming that the liquid  phase thickness is considerably
less than the solid phase thickness, one may use the Dirichlet
condition (\ref{5}). It should be stressed that we neglect the
initial temperature distribution   in solid phase and consider
the process on the stage when two phases have already taken place.

One may claim that formulae  (\ref{1})--(\ref{5}) present the class of BVPs with moving boundaries and take into account  a number of different  situations, which occur in the melting and evaporation processes. For example, setting $Q(t,T_ {1})=\mbox{const}, V_1= \mbox{const} \Phi(t)$ and $ H(t,T_ {1})= \Phi(t)T_ {1}$, where $\Phi(t)$ is a correctly-specified function, one obtains the problem, which is the most typical \cite{ch93}, \cite{ch-od90}. In the case of the process when surfaces are exposed to very  powerful periodic laser impulses these functions take a complicated form \cite{ch-od91}.

The paper is organized as follows. In  section 2, we discuss the existing definition of Lie invariance for BVPs  and present its generalization to a wider class of BVPs. In the next section, we apply the definition derived to BVP (\ref{1})--(\ref{5}). In section 3,  all possible Lie operators of the nonlinear system (\ref{1})--(\ref{2}), which allow us to  reduce the
problem in question to one  for  ordinary differential equations, are found. Finally, we present conclusions in the last section.

\section{Definition of Lie symmetry for BVPs}

We start from the well-known  definition of invariance of a BVP under the given
infinitesimal operator presented in \cite{b-k}. We restrict
ourselves to the case when the basic equation of BVP is a
two-dimensional evolution PDE of $k$th--order ($k\geq 2$). In
this case the relevant BVP may be formulated as follows:
\begin{equation}
u_t=F\left(x, u, u_x, \ldots, u_{x}^{(k)}\right), \ (t,x) \in \Omega
\subset {\bf R}^2 \label{8}
\end{equation}
\begin{equation}
s_a(t,x)=0: \ B_a \left(t,x, u, u_x, \ldots, u_{x}^{(k-1)}\right) =
0, \ a = 1, 2, \ldots, p,\label{9}
\end{equation}
where $F$ and  $B_a$ are smooth functions in the corresponding
domains,   $\Omega$  is    a domain with smooth
boundaries and  $s_a(t,x)$  are  smooth curves. Hereafter the subscripts
$t$ and $x$ denote differentiation with respect to these variables,
$u_{x}^{(j)} = \frac{\partial^j u}{\partial x^j}, j = 1,2,
\ldots,k$. We assume that BVP (\ref{8})--(\ref{9}) has a classical
solution (in a usual sense).

 Consider the infinitesimal generator
\begin{equation}
X = \xi^0 (t,x)\frac{\partial}{\partial t}+\xi^1
(t,x)\frac{\partial}{\partial x} + \eta
(t,x,u)\frac{\partial}{\partial u},\label{10}
\end{equation}
(hereafter  $\xi^0, \xi^1 $ and $\eta$ are known smooth functions),
which defines a Lie symmetry acting on both $(t,x,u)$--space as well
as on its projection to $(t,x)$--space. Let $X^{(k)}$ be the
$k$th--prolongation of the generator $X$ calculated by the
well-known prolongation formulae (see, e.g. \cite{fss}--\cite{ovs}).

\begin{definition}\cite{b-k} The Lie symmetry $X$
 (\ref{10}) is admitted by the boundary value problem
(\ref{8})--(\ref{9}) if and only if:
\begin{itemize}
\item[(a)] $ X^{(k)} \left(F\left(x, u, u_x, \ldots , u_{x}^{(k)}\right)
-u_t \right)=  0 $ when $u$ satisfies (\ref{8});
\item[(b)] $X (s_a(t,x)) = 0$ when $s_a(t,x) = 0, \ a = 1,2, \ldots,p$;
\item[(c)] $ X^{(k-1)} \left(B_a \left(t,x, u, u_x, \ldots ,
u_{x}^{(k-1)}\right) \right) =  0 $ when $B_a \vert_{ s_a(t,x) = 0}=
0$, \ $a = 1,2, \ldots,p$.
\end{itemize}
\end{definition}

One easily  notes that Definition 1 can not  be applied to BVP
(\ref{1})--(\ref{5}), because, firstly, there are  two basic
equations (\ref{1})--(\ref{2}) instead of one,  secondly, the
boundary condition (\ref{5}) is defined on the non-regular manifold  $ x = \infty $, thirdly,
there are several boundary conditions on the moving surfaces
$S_1(t,x)$ and $S_2(t,x)$. Thus, this definition should be extended
to the wider class of BVPs.

Consider a BVP for the  $m$-component  system of  evolution equations
with  two  independent $(t, x)$ and $m$ dependent $u = (u_1, u_2,
\ldots, u_m)$ variables. Let us assume that the $k$th--order ($k\geq
2$) basic equations of evolution type
\begin{equation}
u_t^i=F^i \left(x, u, u_x, \ldots , u_{x}^{(k)}\right), \ i = 1,2,
\ldots, m, \label{14}
\end{equation}
are  defined on a domain $\Omega \subset {\bf R}^2 $ and   there are
three types of boundary conditions, which can   arise in
applications:
\begin{equation}
s_a(t,x)=0: \ B^{j}_a \left(t,x, u, u_x, \ldots ,
u_{x}^{(k_{ja})}\right) = 0,\ a = 1, \ldots, p, \, j =1,\ldots,n_a,
\label{15}
\end{equation}
\begin{equation}
S_b(t,x)=0: \ B^{l}_b \left(t,x, u, \ldots , u_{x}^{(k_{lb})}, S_b,
\frac{\partial S_b}{\partial t}, \frac{\partial S_b}{\partial
x}\right) = 0,\ b = 1, \ldots, q, \, l =1,\ldots,n_b, \label{15a}
\end{equation}
and
\begin{equation}
\gamma_c(t,x)=\infty: \ \Gamma_c \left(t,x, u, u_x, \ldots ,
u_{x}^{({k_{c}})}\right) = 0, \ c = 1, 2, \ldots, r.\label{16}
\end{equation}
Here $k_{ja}< k$ and $k_{lb}< k$ are the given numbers, $s_a(t,x)$
and $\gamma_c(t,x)$ are the known functions, while the functions
$S_b(t,x)$ defining free boundary surfaces must be found. We assume
that all functions arising in (\ref{14})--(\ref{16}) are
sufficiently smooth so that a classical solution exists for this
BVP.

Consider the infinitesimal generator
\begin{equation}
X = \xi^0 (t,x)\frac{\partial}{\partial t}+\xi^1
(t,x)\frac{\partial}{\partial x} + \eta^1
(t,x,u)\frac{\partial}{\partial u^1}+ \ldots +\eta^m
(t,x,u)\frac{\partial}{\partial u^m},\label{17}
\end{equation}
which defines a Lie symmetry acting on both $(t,x,u,S)$--space (here
the notation $S = (S_1, S_2, \ldots, S_q)$ is used) as well as on
its projection to $(t,x)$--space and generates a Lie group of the point
transformations
\begin{equation}
t' = T(t,x,\varepsilon), \ x' = X(t,x,\varepsilon), \ u'_j =
U_j(t,x,u,\varepsilon), \ S'_b = S_b(t,x).\label{18}
\end{equation}

\begin{definition}
BVP (\ref{14})--(\ref{16}) admits the one-parameter Lie group of
 transformations (\ref{18}) generated by   the infinitesimal
operator (\ref{17}) if and only if:
\begin{itemize}
\item[(a)] $ X^{(k)}\left(F^i \left(x, u, u_x, \ldots , u_{x}^{(k)}\right)
-u^i_t\right)=  0$ when the functions $u_i, \, i = 1, 2, \ldots, m$
satisfy (\ref{14}) ;
\item[(b)] $X (s_a(t,x)) = 0$ when $s_a(t,x) = 0,$ \,  $a = 1,2, \ldots,p$;
\item[(c)] $ X^{(k_{ja})}\left(B^{j}_a \left(t,x, u, u_x, \ldots , u_{x}^{(k_{ja})}\right) \right) =  0 $
 when $B^{j}_a \vert_{ s_a(t,x) = 0}= 0,$ \, $a = 1,2, \ldots,p$, $j=1,\ldots,n_a $;
\item[(d)] $ X^{(k_{lb})}\left(B^{l}_b \left(t,x, u, \ldots , u_{x}^{(k_{lb})}, S_b,
\frac{\partial S_b}{\partial t}, \frac{\partial S_b}{\partial
x}\right) \right) =  0 $ when $B^{l}_b \vert_{ S_b(t,x) = 0}= 0$,
 $b = 1, \ldots ,q$, $l=1,\ldots,n_b $;
\item[(e)]  $X_* (\gamma_c^*(\tau,y)) = 0$ when $\gamma_c^*(\tau,y) = 0$, \,$c = 1,2, \ldots,r$;
\item[(f)]  $ X_*^{(k_{c})}\left(\Gamma_c^*\left( \tau, y, u, u_y, \ldots ,
u_{y}^{({k_{c}})}\right)\right) =  0  \,  $
 when $\Gamma_c^* \vert_{ \gamma_c^*(\tau,y) = 0}= 0$, \, $c = 1,2,\ldots,r$,
\end{itemize}

\noindent where $X_*, \, \Gamma_b^*$ and $\gamma_c^*(\tau,y)$ are
operator (\ref{17}),  the functions $\Gamma_b $ and  $
\frac{1}{\gamma_c(t,x)}$, respectively, expressed via the new
independent
variables $\tau=  \left\{ \begin{array}{lll} t, & \mbox{if} & \frac{\partial \gamma_c(t,x)}{\partial x} \not= 0,\\
 \frac{1}{ \gamma_c(t,x)}, & \mbox{if} &  \frac{\partial \gamma_c(t,x)}{\partial x} = 0
\end{array} \right. $  and  $y = $ \\
$ \left\{ \begin{array}{lll} x, & \mbox{if} & \frac{\partial \gamma_c(t,x)}{\partial x} = 0,\\
 \frac{1}{ \gamma_c(t,x)}, & \mbox{if} &  \frac{\partial \gamma_c(t,x)}{\partial x}\not =
 0.
\end{array} \right.$
\end{definition}

It should be noted that this Definition coincides with Definition
1 if $m=1, \ n_a=1$ and there are no boundary conditions of the
form (\ref{15a}) and (\ref{16}). On the other hand, the following
example shows that the generalization of Definition 1 presented
above is non-trivial.

{\bf Example.}
Let us consider BVP (\ref{14})--(\ref{15}), which includes also the
boundary conditions
\[
x = \infty: \ \Gamma_c(u) = 0, \ c = 1, 2, \ldots, r.
\]
Assume that system (\ref{14}) and conditions (\ref{15}) are
invariant under the group of translations on the plane $(t,x)$:
\begin{equation}\label{20}
t' = t + \lambda_1 \varepsilon, \ \ x' = x + \lambda_2\varepsilon, \
\ u' = u, \ \lambda_1 \lambda_2 \neq 0,
\end{equation}
so that  the  corresponding  infinitesimal generator
\begin{equation}\label{21}
X = \lambda_1 \partial_t + \lambda_2 \partial_x
\end{equation}
satisfies items (a)-(c) from Definition 2. Moreover, according to
Definition 2 the operator $X$ of the form (\ref{21}) and the
functions $\Gamma_c(u)$ and  $\gamma_c(t,x)=x$ take the form
\[
X_* = \lambda_1 \partial_{\tau} - \lambda_2 y^2 \partial_y
\]
$\Gamma_c^*(u)=\Gamma_c(u)$ and $\gamma_c^*(\tau,y)=y$,
respectively. Now one easily checks that items (e) and  (f) from
Definition 2 are satisfied. Thus, BVP in question   admits the Lie
group of point transformations (\ref{20}) generated by   the
infinitesimal generator (\ref{21}).

However,  the problem with non-regular manifold $\gamma_c(t,x) =
\infty$ occurs if one generalizes Definition 1 in
 the standard  way and   formulates items (e) and  (f) from Definition 2 like those (b) and (c).
For example, this manifold can not be replaced by the regular one
$\gamma_c(t,x) - L=0$, where $L \rightarrow \infty$ because this
leads to the requirement
\[
\lim \limits_{L \rightarrow \infty} \left. X (\gamma_c(t,x)- L)
\right \vert_{\gamma_c(t,x) = L} = 0.
\]
 On the other hand, this requirement is not satisfied in  the example
 presented above
 because
\[
\lim \limits_{L \rightarrow \infty} \left. X (x - L) \right \vert_{x
= L} = \lim \limits_{L \rightarrow \infty} \lambda_2 = \lambda_2
\neq 0.\] Thus, item (e) in Definition 2 can not be formulated
like item (b).

Hereafter Definition 2 will be applied to derive the invariance
operators of BVP (\ref{1})--(\ref{5}).

\section{Invariance of BVP (\ref{1})--(\ref{5}) under the
Lie generators}

 It should be noted that BVP (\ref{1})--(\ref{5}) can be
simplified if one applies the Goodman  substitution
\begin{equation}\label{2.13}
u = \phi_1(T_1) \equiv \int\limits_0^{T_{1}} {c_{1}(\zeta)
\rho_1(\zeta)}\,d\zeta, \quad v = \phi_2(T_2) \equiv
\int\limits_0^{T_{2}} {c_{2}(\xi) \rho_2(\xi)}\,d\xi.
\end{equation}
Substituting (\ref{2.13}) into  (\ref{1})--(\ref{5}) and making the
relevant calculations, we arrive at the equivalent BVP of the form
\begin{eqnarray}
& & \frac{\partial u}{\partial t}  =  \frac{\partial}{\partial
x}\left(d_{1}(u) \frac{\partial
u}{\partial x}\right),\label{2.14}  \\
 & & \frac{\partial v}{\partial t}  =  \frac{\partial}{\partial
x}\left(d_{2}(v)
\frac{\partial v}{\partial x}\right),\label{2.15} \\
& & \qquad S_{1}(t,x) = 0: \ d_{1}(u)
\frac{\partial u}{\partial x} = \hat{\rho}_1(u)L_v V_1- \hat{q}(t,u), \ V_1 = \hat{h}(t,u),\label{2.16} \\
& & \qquad S_{2}(t,x) = 0: \ d_{2}(v_m) \frac{\partial v}{\partial
x} = d_{1}(u_m) \frac{\partial u}{\partial x} + \hat{\rho}_2(v_m)L_m
V_2, \ u = u_{m},  v =  v_{m},\label{2.17}
\\ & & \qquad x  =  +\infty: \ v = v_{\infty},\label{2.18}
\end{eqnarray}
where $u_m = \int\limits_0^{T_{m}} {c_{1}(\zeta)
\rho_1(\zeta)}\,d\zeta$, $v_m = \int\limits_0^{T_{m}} {c_{2}(\xi)
\rho_2(\xi)}\,d\xi$, $v_{\infty} = \int\limits_0^{T_{0}} {c_{2}(\xi)
\rho_2(\xi)}\,d\xi$; $d_1(u) =
\frac{\lambda_1(\phi^{-1}_1(u))}{c_1(\phi^{-1}_1(u))\rho_1(\phi^{-1}_1(u))}$,
$d_2(v) =
\frac{\lambda_2(\phi^{-1}_2(v))}{c_2(\phi^{-1}_2(v))\rho_2(\phi^{-1}_2(v))}$;
$\hat{\rho}_1(u) = \rho_1(\phi^{-1}_1(u))$, $\hat{\rho}_2(v) =
\rho_2(\phi^{-1}_2(v))$, $\hat{q}(t,u) = Q(t,\phi^{-1}_1(u))$, and
$\hat{h}(t,u) = H(t,\phi^{-1}_1(u))$ (here $\phi^{-1}_i$ are inverse
functions to $\phi_i$,  the functions $d_1(u)$ and $d_2(v)$ are
strictly  positive and $v_m \neq v_{\infty}$).

Now one sees that BVP (\ref{2.14})--(\ref{2.18}) is based on the
standard nonlinear heat equations (NHE). Lie symmetries of the
non-coupled system (\ref{2.14})--(\ref{2.15}) can be easily derived
from paper \cite{ch-king4}, where reaction-diffusion systems of more
general form have been investigated. The result is presented in
Table 1 (note, we do not consider the case when this system is
linear). Using Definition 2 and Lie symmetries from Table 1 the
following theorem can be established.

\begin{theorem} A nonlinear BVP of the form (\ref{2.14})--(\ref{2.18}) admits a Lie symmetry operator of system (\ref{2.14})--(\ref{2.15}) if and only if the operator in question up to the local
transformations $x\rightarrow x+x_{0}, \ t\rightarrow t+t_{0} (x_{0}
\in {\bf R}, \ t_{0} \in {\bf R})$ is equivalent either to the
operator
\begin{equation}
X_{1} = \partial_{t} + \mu \partial_{x}, \ \mu \in {\bf R}
\label{2.21}
\end{equation}
 or to
\begin{equation}
X_{2} = 2t\partial_{t} + x\partial_{x}.\label{2.22}
\end{equation}
Moreover, the functions $\hat{f}(t,u), \ \hat{q}(t,u)$
 must  have the correctly specified forms
  \begin{equation}
\hat{h}(t,u) = h(u), \ \hat{q}(t,u)=q(u)\label{2.23}
\end{equation}
and
 \begin{equation}
\hat{h}(t,u) = \frac{h(u)}{\sqrt{t}}, \ \hat{q}(t,u) =
\frac{q(u)}{\sqrt{t}},\label{2.24}
\end{equation}
respectively. Here $q(u)$ and $h(u)$ are arbitrary smooth functions.
\end{theorem}

{\bf Proof.} Firstly, we derive the group classification of the NHE system
(\ref{2.14})--(\ref{2.15}) using the determining equations presented
in \cite{ch-king4}.
 If $d_1(u)$ and $d_2(v)$ are
arbitrary functions then those equations immediately give the
three-dimensional maximal algebra of invariance (MAI) $A = \langle
\partial_{t},
\partial_{x}, 2t \partial_{t} + x
\partial_{x}\rangle$ called the  principal algebra. There are five special cases when
an  extension of  the principal algebra occurs and they are listed
in Table 1.

It should be stressed that each NHE system admitting four- or
five-dimensional Lie algebra is reduced to one of those with
diffusivities  from Table 1 by the equivalence transformations
\begin{equation}\label{2.25}
t \rightarrow  e_0t + t_{0},\quad x \rightarrow e_1x + x_{0}, \quad
u \rightarrow  e_2u + u_{0}, \quad v \rightarrow  e_3v + v_{0},
\end{equation}
where $e_i \not=0 \ (i = 0, \ldots,3), t_{0}, x_{0}, u_0$, and
$v_{0}$ are arbitrary parameters. It turns out  that the class of
BVPs (\ref{2.14})--(\ref{2.18}) is also invariant  under
transformations (\ref{2.25}) and this can be easily checked by
direct calculations.

Thus, on the next stage,  we need to apply
Definition 2 only to BVP (\ref{2.14})--(\ref{2.18}) in six  cases
listed in Table 1.

\textbf{Table 1.} {\bf  Lie algebras of the NHE system
(\ref{2.14})--(\ref{2.15}).}
 {\renewcommand{\arraystretch}{1.5}
\begin{center}
\begin{tabular}{|c|c|c|c|}
  \hline
  \emph{ no} & \emph{$d_1(u)$} & \emph{$d_2(v)$} & \emph{Basic operators of MAI}\\
  \hline\hline
1. & $\forall$ & $\forall$ & $ A = \langle
\partial_{t},
\partial_{x}, 2t \partial_{t} + x
\partial_{x}\rangle$\\
  2. & $\forall$ & 1 & $ A, v \partial_v, \beta(t,x) \partial_v $\\
  3. & $e^u$ & $e^v$ & $ A, x \partial_{x} + 2 \partial_{u} +  2 \partial_{v}$  \\
  4. & $e^u$ & $v^m$& $ A, x \partial_{x} + 2 \partial_u + \frac{2}{m} v \partial_{v} $ \\
  5. & $u^n$ & $v^m$& $ A, x \partial_{x} + \frac{2}{n} u \partial_u + \frac{2}{m} v \partial_{v} $ \\
  6. & $u^{-\frac{4}{3}}$ & $v^{-\frac{4}{3}}$ & $ A, x \partial_{x} - \frac{3}{2} u \partial_u - \frac{3}{2} v
  \partial_{v}, x^2 \partial_x - 3 x u \partial _u - 3 x v \partial_v $\\
  \hline
\end{tabular}
\end{center}}

Let us consider the first case when  the  functions $d_1(u)$ and
$d_2(v)$ are arbitrary. In this case the most general form of the
Lie symmetry generator is
\begin{equation}\label{2.26}
X = (\lambda_{1} +
2\lambda_{3}t)\partial_{t}+(\lambda_{2}+\lambda_{3}x)\partial_{x}
\end{equation}
(hereinafter  $\lambda$ with indices are arbitrary constants).
 Obviously, operator (\ref{2.26})  can be reduced
either to the form (\ref{2.21}) (if $\lambda_{3}=0$ and
$\lambda_{1}\not=0$) or to (\ref{2.22}) (if $\lambda_{3}\not=0$) by
the local transformations
 \[ x\rightarrow x+x_{0}, \ t\rightarrow
t+t_{0}\left( x_{0}\in {\bf R}, \ t_{0} \in {\bf R} \right),
\]
which belong to  (\ref{2.25}).

Note that the case $\lambda_{3}=\lambda_{1}=0$
leads to the  invariance operator  $X=\partial_{x}$. However, this operator generates
    the Lie ansatz, which  does not alow existing  the moving boundary
surfaces $S_k(t,x), \ k = 1,2$, hence, we do not consider this non-physical  case.

Let us apply Definition 2 to prove the invariance of BVP
(\ref{2.14})--(\ref{2.18}) under the Lie symmetry operator
(\ref{2.21}). Applying the first prolongation of generator
(\ref{2.21}), $X_1^{(1)} = X_1$, to  the  boundary conditions
(\ref{2.16}), we obtain
\begin{eqnarray}
& & X_1\left. \left(d_{1}(u) \frac{\partial u}{\partial x} -
\hat{\rho}_1(u)L_v V_1 + \hat{q}(t,u) \right)\right \vert_{\cal M} =
\frac{\partial \hat{q}(t,u)}{\partial t},\nonumber  \\ & & X_1
\left. \left(V_1 - \hat{h}(t,u)\right)\right \vert_{\cal N} =
-\frac{\partial \hat{h}(t,u)}{\partial t}\nonumber.
\end{eqnarray}
where  the  manifold ${\cal M} = \left \{S_1(t,x) = 0: d_{1}(u)
\frac{\partial u}{\partial x} = \hat{\rho}_1(u)L_v V_1 -
\hat{q}(t,u)\right \}$ and the manifold ${\cal N} = $ \\ $\left
\{S_1(t,x) = 0: V_1 = \hat{h}(t,u) \right \}$. According to item (d)
from Definition 2 we arrive at the restriction on the function
$\hat{q}(t,u)$ and $\hat{h}(t,u)$
\begin{eqnarray}
& & \frac{\partial \hat{q}(t,u)}{\partial t} = 0 \Leftrightarrow
\hat{q} = q(u),\nonumber \\ & & \frac{\partial
\hat{h}(t,u)}{\partial t} = 0 \Leftrightarrow \hat{h} = h(u)
\nonumber.
\end{eqnarray}
The invariance of conditions (\ref{2.17}) are trivially fulfilled
because they do not involve the variables $t$ and $x$ in explicit
form. Finally, condition (\ref{2.18}) satisfies items (e) and (f)
from Definition 2 because this immediately follows from Example.

The invariance of BVP (\ref{2.14})--(\ref{2.18}) under generator
(\ref{2.22}) is checked in a similar way. Note that
 the first prolongation of generator (\ref{2.22}) has the more
complicated form than one of (\ref{2.21}), namely
\begin{equation}\label{46}
X_2^{(1)} = 2t \partial_t + x \partial_x - 2u_t \partial_{u_t} -
2v_t \partial_{v_t} - u_x \partial_{u_x} - v_x \partial_{v_x} -
2S_t^1 \partial_{S_t^1} - 2S_t^2 \partial_{S_t^2} - S_x^1
\partial_{S_x^1} - S_x^2 \partial_{S_x^2}
\end{equation}
Since the boundary conditions (\ref{2.16}) involve the first-order
derivatives (we remind that $V_k = - \frac{\partial S_k}{\partial t} /
\frac{\partial S_k}{\partial x}$)
  we have calculated  how
 operator  (\ref{46}) acts on these boundary
conditions according to Definition 2  and obtained two ODEs of the
form
\[ 2t\frac{\partial \hat{q}(t,u)}{\partial t}+\hat{q}(t,u)=0, \quad
2t\frac{\partial \hat{h}(t,u)}{\partial t}+\hat{h}(t,u)=0, \]
which lead to the  restrictions (\ref{2.24}).

Thus, the case of arbitrary functions $d_1(u)$ and $d_2(v)$ is
completely examined.

It turns out that any other Lie generator arising in Table 1 does
not satisfy Definition 2 for BVP (\ref{2.14})--(\ref{2.18}). Let
us consider, for instance, case 4 of Table 1. Here, the most general
form of the Lie symmetry generator is
\begin{equation}\label{47}
X_3 = (\lambda_{1} +
2\lambda_{3}t)\partial_{t}+(\lambda_{2}+(\lambda_{3}+
\lambda_{4})x)\partial_{x} + 2\lambda_{4} \partial_{u} +
\frac{2}{m}\lambda_{4} v \partial_{v},
\end{equation}
where  we  assume $\lambda_4 \not = 0$, otherwise $X_3 $ takes  form
(\ref{2.26}).

In order to be invariant under generator (\ref{47}), the second
condition from (\ref{2.17}) must satisfy item (d) from Definition
2
\[
X_3 \left.(u - u_m)\right\vert_{\cal P} = 0,
\]
where ${\cal P} = \left \{S_2(t,x) = 0: u = u_m \right\}$, however,
this leads to the restriction $ \lambda_4 = 0$. Thus,  the contradiction
is obtained and we conclude that generator (\ref{47}) is not Lie
symmetry of any BVP (\ref{2.14})--(\ref{2.18}) with diffusivities
from case 4 of Table 1.

In a similar way, cases 2,3, 5 and 6 listed in  Table 1 have been
examined and contradictions with boundary value conditions
established.

The proof is now complete.$\blacksquare$

\begin{remark} The group classification of the NHE system (\ref{2.14})--(\ref{2.15}) presented in Table 1 takes into account the trivial
discrete transformations $u \rightarrow v, \, v \rightarrow u $
admitted by this system. Nevertheless, the class of BVPs
(\ref{2.14})--(\ref{2.18}) is not invariant under these
transformations (see conditions (\ref{2.16})) it does not affect
Theorem 1.
\end{remark}

Using Theorem 1  one can reduce the class of  BVPs
(\ref{2.14})--(\ref{2.18}) to two classes  of  BVPs for second-order
ODEs.   In fact, substituting  ans\"atze
\begin{eqnarray}
& & u = u(\xi), \ \ v = v(\xi), \ \ S_k = S_k(\xi), \ \ \xi = x - \mu t, \label{51} \\
& & u = u(\omega), \ \ v = v(\omega), \ \ S_k = S_k(\omega), \ \
\omega = \frac{x}{\sqrt{t}}, \label{52}
\end{eqnarray}
generated by operators    (\ref{2.21}) and (\ref{2.22}),
respectively, and taking into account restrictions (\ref{2.23}) and
(\ref{2.24}) the following consequences are obtained.

\begin{corollary}  Ansatz (\ref{51}) reduces
a nonlinear BVP of the form (\ref{2.14})--(\ref{2.18}) with the
coefficient restrictions (\ref{2.23}) to the following BVP for the
second-order ODEs
\begin{eqnarray}
& & \frac{d}{d \xi}\left(d_{1}(u)
\frac{d u}{d \xi}\right) + \mu \frac{d u}{d \xi} = 0, \ \ \ 0 < \xi < \delta,  \label{57}\\
 & & \frac{d}{d \xi}\left(d_{2}(v)
\frac{d v}{d \xi}\right) + \mu \frac{d v}{d \xi} = 0, \ \ \ \xi > \delta, \label{58}\\
& & \qquad \xi = 0: \ d_{1}(u)
\frac{d u}{d \xi} = \hat{\rho}_1(u)L_v \mu - q(u), \ \mu = h(u), \label{59}\\
& & \qquad \xi = \delta: \ d_{2}(v_m) \frac{d v}{d \xi} = d_{1}(u_m)
\frac{d u}{d \xi} + \hat{\rho}_2(v_m)L_m \mu, \ u = u_{m}, \ v =
v_{m}, \label{60}\\
& & \qquad \xi  =  +\infty: \ v = v_{\infty}, \label{61}
\end{eqnarray}
where $\delta$ and $\mu$ are parameters to be found.
\end{corollary}

\begin{corollary}
Ansatz (\ref{52}) reduces a nonlinear BVP of the form
(\ref{2.14})--(\ref{2.18}) with the coefficient restrictions
(\ref{2.24}) to the following BVP for the second-order ODEs
\begin{eqnarray}
& & \frac{d}{d \omega} \left(d_{1}(u) \frac{d u}{d \omega}\right) + \frac{\omega}{2} \frac{d u}{d \omega} = 0,
\ \ \ \omega_1 < \omega < \omega_2, \label{62}\\
& & \frac{d}{d \omega} \left(d_{2}(v) \frac{d v}{d \omega}\right) + \frac{\omega}{2} \frac{d v}{d \omega} = 0,
\ \ \ \omega > \omega_2, \label{63} \\
&& \qquad \omega = \omega_{1}: d_{1}(u) \frac{d u}{d \omega} =
\hat{\rho}_1(u) L_v \frac{\omega_{1}}{2} - q (u), \ \frac{\omega_1}{2} = h(u), \label{64}\\
& & \qquad \omega = \omega_{2}: d_{2}(v_m) \frac{d v}{d \omega} =
d_{1}(u_m) \frac{d u}{d \omega} + \hat{\rho}_2(v_m)L_m \frac{\omega_{2}}{2},\ u = u_{m}, \ v = v_{m},\label{65} \\
& & \qquad \omega = + \infty: v = v_{\infty},\label{66}
\end{eqnarray}
where $\omega_1$ and $\omega_2$ are  parameters to be found.
\end{corollary}

It should be noted that each BVP
 (\ref{57})--(\ref{61})
with the correctly specified functions  $q(u)$ and $h(u)$ arising in
(\ref{59}) can exactly be solved and an exact solution can be
presented at least in an implicit form (see for details
\cite{ch93}). Some BVPs
 (\ref{62})--(\ref{66}) with the
correctly specified coefficients  arising in  the problem were
exactly solved in particular cases  \cite{ch-kov-09}.

\section{Conclusions}
In this paper, the class of (1+1)--dimensional nonlinear boundary
value problems (\ref{1})--(\ref{5}), modeling the  process of
melting and evaporation of metals, is studied  by means of the
classical Lie symmetry method. New definition of invariance in Lie's
sense
is presented. This
definition is applicable to  wide classes  of BVPs, including those
with several basic equations, with moving boundaries, and with
boundary conditions on non-regular manifolds.

Theorem 1 giving all possible  operators of Lie's invariance for
the class of  BVPs (\ref{1})--(\ref{5}) is proved and the
corresponding  corollaries, which allow us to reduce the problems in
question to those for ODE systems, are also obtained.

Finally, we note that Theorem 2 from the recent paper
\cite{ch-kov-09} follows as a particular case from Theorem 1,
nevertheless  Definition 2  was not used in \cite{ch-kov-09} because
the direct application of Lie point transformations was applied
therein. This means that definition of Lie's invariance for BVPs can
be derived in different ways. We are going to discuss this in
details in a forthcoming paper.


\end{document}